\documentclass[conference,10pt]{IEEEtran}
\normalsize
\usepackage{bbm}
\usepackage{adjustbox}
\usepackage{enumitem}
\usepackage{cite,algorithm,algorithmic,amsmath,amssymb,amsthm,empheq,mhsetup}
\usepackage{subfigure,amsfonts,balance}
\usepackage{epstopdf}
\usepackage{enumitem}
\usepackage{setspace}
\usepackage[dvipsnames]{xcolor}
\DeclareMathOperator{\EEE}{\mathbb{E}}

\DeclareMathOperator{\f}{\pmb{f}}

\DeclareMathOperator{\FF}{\mathcal{F}}

\DeclareMathOperator{\K}{\mathcal{K}}

\DeclareMathOperator{\vv}{\pmb{v}}

\DeclareMathOperator{\CN}{\mathcal{CN}}

\DeclareMathOperator{\MM}{\mathcal{M}}

\DeclareMathOperator{\NN}{\mathcal{N}}

\DeclareMathOperator{\rr}{\pmb{r}}
\DeclareMathOperator{\x}{\pmb{x}}

\DeclareMathOperator{\uu}{\pmb{u}}

\DeclareMathOperator{\q}{\pmb{q}}
\DeclareMathOperator{\g}{\pmb{g}}

\DeclareMathOperator{\ETA}{\pmb{\eta}}

\DeclareMathOperator{\ZETA}{\pmb{\zeta}}

\newtheorem{remark}{Remark}

\ifCLASSINFOpdf
\else
\fi

\begin{document}
\fontsize{10}{12}\rm
%
\title{\Huge How Does Cell-Free Massive MIMO Support Multiple Federated Learning Groups?\vspace{-2mm}}
\author{
\IEEEauthorblockN{
Tung T. Vu\IEEEauthorrefmark{1},
Hien Quoc Ngo\IEEEauthorrefmark{1},
Thomas L. Marzetta\IEEEauthorrefmark{2}, and
Michail Matthaiou\IEEEauthorrefmark{1}
}
\IEEEauthorblockA{\small\IEEEauthorrefmark{1}Institute of Electronics, Communications, and Information Technology (ECIT), Queen's University Belfast, Belfast BT3 9DT, UK}
\IEEEauthorblockA{\small\IEEEauthorrefmark{2}Department of Electrical and Computer Engineering, New York University, 11201 Brooklyn, NY}
\IEEEauthorblockA{
Email: t.vu@qub.ac.uk, hien.ngo@qub.ac.uk, tom.marzetta@nyu.edu, m.matthaiou@qub.ac.uk
}
\vspace{-9mm}
}

\maketitle
\allowdisplaybreaks
\vspace{-0mm}
\begin{spacing}{1}
\begin{abstract}
Federated learning (FL) has been considered as a promising learning framework for future machine learning systems due to its privacy preservation and communication efficiency. In beyond-5G/6G systems, it is likely to have multiple FL groups with different learning purposes. This scenario leads to a question: How does a wireless network support multiple FL groups? As an answer, we first propose to use a cell-free massive multiple-input multiple-output (MIMO) network to guarantee the stable operation of multiple FL processes by letting the iterations of these FL processes be executed together within a large-scale coherence time. We then develop a novel scheme that asynchronously executes the iterations of FL processes under multicasting downlink and conventional uplink transmission protocols. Finally, we propose a simple/low-complexity resource allocation algorithm which optimally chooses the power and computation resources to minimize the execution time of each iteration of each FL process. 
\end{abstract}
\end{spacing}


%
\IEEEpeerreviewmaketitle

\vspace{-2mm}
\section{Introduction}
\vspace{-0mm}
\label{sec:Introd}
The concept of federated learning (FL), which was first introduced in \cite{mcmahan17AISTATS}, proposed the radical idea of ``not sending raw data to third party companies during learning processes''. It is definitely one shot with multiple critical goals including privacy preservation and communication efficiency. Since then, FL has quickly become  an innovation trend for digital systems with a wide range of applications, such as healthcare and self-driving cars to name but a few \cite{chen20IS,niknam20CM}. 
In FL, many users (UEs) cooperate to implement the learning. More specifically, first, each UE collects the data and trains a learning model locally. This local learning model will be sent to the central server. The central server then uses all local training models from all UEs to compute the global update, which is then sent back to the UEs for their further local training updates. This process is done iteratively until a certain learning accuracy level is achieved. In many scenarios, the above iterative process needs to be implemented over a wireless network. Thus, designing a good wireless framework to support FL is of particular importance. 

There are several works on the implementation of FL over wireless networks, such as \cite{chen21TWC,amiri20TWC,vu20TWC} and references therein. These works can be classified into learning-oriented and communication-oriented categories. The first category of papers seeks to improve learning performance (i.e., test accuracy) by reducing the detrimental impacts of wireless networks, such as thermal noise, fading, and estimation errors, on FL \cite{chen21TWC,amiri20TWC}. 
The second category contributes to reducing the time (in sections) of an FL process executed over wireless networks by optimally designing communication schemes \cite{vu20TWC}. However, all above works considered the case of a single FL group.

It is foreseen that future wireless FL systems will include multiple FL groups, where multiple groups of UEs with different learning purposes participate in multiple FL processes and get the learning results within a short period of time. To support multiple FL groups, a wireless network needs to simultaneously  provide very high quality of service (high data rate and high communication reliability) to all UEs in all FL groups. This is a very challenging exercise and requires a suitable wireless communication framework. To the best of our knowledge, there has not been any work studying wireless networks supporting multiple FL groups in the prior literature.

\vspace{-0mm}
\textit{Paper Contribution:} Following the communication-oriented paradigm, this work proposes a novel communication scheme to support multiple FL groups. We first propose using cell-free massive MIMO (CFmMIMO) and let multiple iterations (each corresponding to a FL process of a group) be executed in one large-scale coherence time.\footnote{Large-scale coherence time is defined as a time interval where the large-scale fading coefficient remains reasonably invariant.} Thanks to the high array gain, multiplexing gain, and macro-diversity gain, CFmMIMO can provide very good quality of services for all users in the area of interest \cite{ngo17TWC}, and hence, it is expected to guarantee a stable operation of each iteration and then the whole FL process. A specific transmission protocol is proposed where the steps within one FL iteration, i.e., downlink transmission, computation at UEs' devices, uplink transmission, are asynchronous.
The downlink follows the multicasting protocol, while the uplink follows the conventional multiuser transmission. We then propose an algorithm to allocate transmit power and processing frequency to reduce the execution time of each FL iteration of each FL group. Numerical results show that our proposed scheme significantly reduces  the FL execution time compared to baseline schemes, and confirm that CFmMIMO is a better choice than colocated massive MIMO with the same maximum ratio technique for supporting multiple FL groups.
\vspace{-3mm}

\section{Proposed Scheme and System Model}
\label{sec:SystemModel}
\vspace{-1mm}
\subsection{Multiple Federated Learning Framework}\label{sec:FLframework}
\vspace{-1mm}
We consider a multiple FL system, where a central server supports $N$ FL groups. The $n$-th group  has $K_n$ single-antenna UEs.
We assume that each UE participates only in one FL group. The FL frameworks of all groups can be different in terms of loss functions but share the same four steps in each iteration as follows \cite{mcmahan17AISTATS,tran19INFOCOM}: 
\vspace{-1mm}
\begin{enumerate}[label={(S\arabic*)}]
\item A central server sends a global update to all the UEs of each group.
\item With the received global update, each UE updates and solves its local learning problem over its local data and then computes its local update.
\item Each UE sends its computed local update to the central server.
\item The central server computes the global update by aggregating the received local updates.
\end{enumerate}
\vspace{-1mm}
The above process is implemented iteratively until a certain learning accuracy level is achieved.
\vspace{-2mm}

\subsection{CFmMIMO-based Multiple-FL System Model}
\vspace{-1mm}
To support multiple learnings discussed in Section~\ref{sec:FLframework}, we propose to use CFmMIMO, i.e. Steps (S1) and (S3) of each learning iteration can be done via the downlink and the uplink of a CFmMIMO system, respectively. Our proposed CFmMIMO-based multiple-FL system includes $M$ single-antenna APs simultaneously serving $N$ FL groups in the same frequency bands under time-division-duplexing operation. All the APs are connected to the central processing unit (CPU) (i.e., the central server) via high-capacity backhaul links, and thus, the transmission times between the CPU and all the APs are negligible. 

\begin{figure}[t!]
\centering
\includegraphics[width=0.46\textwidth]{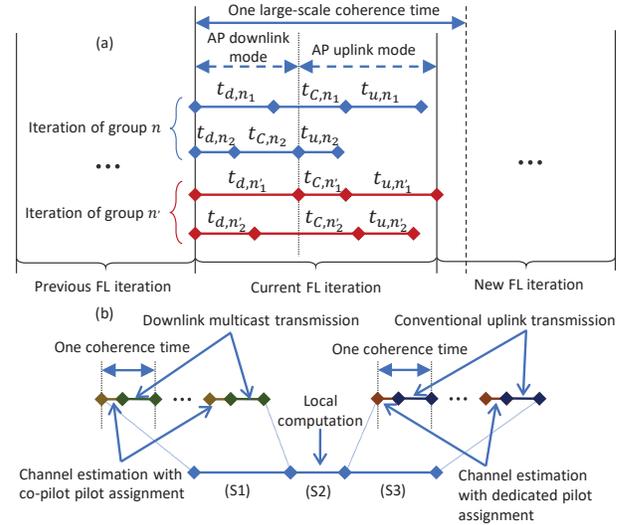}
\vspace{-5mm}
\caption{(a): Illustration of FL iterations over the considered CFmMIMO network with two groups $n,n'$ and two UEs for each group. (b): Detailed operation of one FL iteration of each group.}
\label{fig:time}
\end{figure}

We assume that each FL iteration of each FL group is executed within a large-scale coherence time. This assumption is reasonable since in many scenarios, the execution time of one FL iteration is smaller than the large-scale coherence time \cite{vu20TWC}.
With this assumption, we then propose a specific transmission protocol to support the learnings of $N$ FL groups for each FL iteration as shown in Fig.~\ref{fig:time} (a). 
First, all groups start their FL iterations at the same time when the APs switch to a downlink mode. During this mode, all APs simultaneously send the global updates from the CPU to all users in all groups (corresponding to Step~(S1)). Each user will start its local computation if it receives enough global training update (corresponding to Step~(S2)). Then, the APs switch to an uplink mode immediately after the receptions of the global training update are completed at all the UEs. During this mode, the users will send  their computed local update to the APs (and hence, the CPU) (corresponding to Step~(S3)) if they finish the local computation. Since the APs cannot transmit and receive at the same time, the UEs that already finished their uplink transmission will have to wait until all other UEs complete their transmission to start their new FL iterations. 

The global and local updates in Steps (S1) and (S3) can be transmitted in one or multiple (small-scale) coherence times based on their sizes, as shown in Fig.~\ref{fig:time}(b). Each coherence block in Step (S1) (or (S3)) includes the channel estimation phase and the downlink (or uplink) payload data phase. 

Assume that at a particular time, there are $N$ iterations of $N$ FL groups being served. 
Let $\MM\triangleq\{1,\dots,M\}$, $\NN\triangleq\{1,\dots,N\}$, and $\K_n$ be the sets of APs, groups, and the indices of the UEs in group $n$, respectively. In the following, we will present the details of each step.


\subsubsection{Step (S1)}
In this step, the CPU wants to send the global updates to all users in all groups. This can be done through the downlink transmission. Since the UEs of each group receive the same global update, we propose to use a multicast transmission with a co-pilot assignment for channel estimation. Each coherence block of this step includes two main phases: uplink channel estimation and downlink multicasting.

\textbf{Uplink channel estimation}:
In each coherence block of length $\tau_c$ symbols, all the UEs of a given group send the same pilot of length $\tau_{cp}$ symbols to the CPU \cite{doan17CL}. We assume that the pilots of $N$ groups are pairwisely orthogonal, which requires $\tau_{cp}\geq N$. 
Denote by $g_{mn_k} \!=\! (\beta_{mn_k})^{1/2}\tilde{g}_{mn_k}$ the channel from UE $k$ of group $n$ to AP $m$, where $\beta_{mn_k}$ and $\tilde{g}_{mn_k} \sim \CN(0,1)$ are the large-scale fading and small-scale fading coefficients, respectively. At AP $m$, $g_{mn_k}$ is estimated by using the received pilot signals together with the minimum mean-square error (MMSE) estimation technique. By following \cite{doan17CL}, we can obtain the MMSE estimate  of $g_{mn_k}$ as $\hat{g}_{mn_k}$ which is 
distributed according to $\CN(0,\hat{\sigma}_{mn_k}^2)$, where
$\hat{\sigma}_{mn_k}^2= \frac{\tau_{cp} \rho_{p} (\beta_{mn_k})^2 }{ \tau_{cp} \rho_{p} \sum_{\ell\in\K_n}  \beta_{mn_\ell} + 1}$.


\textbf{Downlink multicasting}:
Since the global training updates for all users in a given group are the same, the CPU encodes the global training update intended for the UEs of group $n$ into the same symbol $s_{d,n}$ and sends all symbols $s_{d,n}, \forall n\!\!\!\in\!\!\!\NN$, to all the APs, where $\EEE\{|s_{d,n}|^2\}\!\!=\!\!1$ and $\EEE\{\}$ is the expectation operator.
After receiving the symbols from the CPU, the APs use conjugate beamforming to precode these symbols. The transmitted signal at AP $m$ is thus given as $x_{d,m}\!\!=\!\! \sqrt{\rho_{d}}\!\sum_{n'\in\NN}\!\sum_{n'_\ell\in\K_{n'}}\!\!\sqrt{\eta_{mn'}}\hat{g}_{mn'_\ell}^*s_{d,n'}$,
 where $\rho_{d}$ is the maximum normalized transmit power at each AP and $\eta_{mn'}, \forall m\!\in\!\MM,n'\!\in\!\NN$, is a power control coefficient associated with AP $m$ and FL group $n$. The transmitted power at AP $m$ is required to meet the average normalized power constraint, i.e., $\EEE\{|x_{d,m}|^2\}\!\leq \!\rho_d$,
which can be expressed as the following per-AP power constraint:
\begin{align}
\label{powerdupperbound}
\sum_{n\in\NN}\sum_{n_k\in\K_n}\sigma_{mn_k}^2\eta_{mn} \leq 1, \forall m\in\MM.
\end{align}

\begin{figure*}[t!]
\begin{align}\label{Rdmulti}
R_{d,n_k}(\ETA)=\frac{\tau_c-\tau_{cp}}{\tau_c}B\log_2\!\Bigg(\!1+\!
\frac
{\rho_d\big(\sum_{n_\ell\in\K_n}\sum_{m\in\MM}\eta_{mn}^{1/2}\hat{\sigma}_{mn_k} \hat{\sigma}_{mn_\ell}\big)^2}
{\rho_d \sum_{n'\in\NN}  \sum_{m\in\MM} \eta_{mn'} \beta_{mn_k} \big( \sum_{n'_\ell\in\K_{n'}}  \hat{\sigma}_{mn'_\ell} \big)^2  
+1}\Bigg),
\end{align}
\hrulefill
\vspace{-7mm}
\end{figure*}


By following \cite[Proposition~1]{zhang20WCNC}, we can obtain the achievable rate $R_{d,n_k}(\ETA)$ for UE $k$ of group $n$ given in \eqref{Rdmulti} shown at the top of the next page, where $\ETA\triangleq\{\eta_{mn_k}\}_{m\in\MM,n\in\NN,k\in\K_n}$, and $B$ is the bandwidth.


\textbf{Downlink delay}:
Let $S_{d,n}$ (bits) be the data size of the global training update of group $n$. 
The transmission time from the APs to UE $k$ in group $n$ is given by 
\begin{align}\label{}
t_{d,n_k}(\ETA) = \frac{S_{d,n}}{R_{d,n_k}(\ETA)}. 
\end{align}
\vspace{-3mm}

\subsubsection{Step (S2)}
In this step, after receiving the global update from the APs in Step (S1), all the UEs run the same number of local computation $L$ over their local data sets to compute local updates.

\textbf{Computation delay}: 
 Denote by $c_{n_k}$ (cycles/sample) the number of processing cycles for a UE $k$ to process one data sample, which is known \emph{a priori} by an offline measurement \cite{tran19INFOCOM}. Let $D_{n}$ (samples) and $f_{n_k}$ (cycles/s) be the size of the local data set and the processing frequency of the UE $k$ of group $n$, respectively. The time for local computation at UE $k$ of group $n$ is then given by \cite{vu20TWC,tran19INFOCOM}
\begin{align}\label{}
t_{C,n_k}(f_{n_k}) = \frac{LD_nc_{n_k}}{f_{n_k}}.
\end{align}
\vspace{-3mm}



\subsubsection{Step (S3)}
In this step, each UE sends its local update to the CPU. This is done via the uplink transmission which include two main phases (uplink channel estimation and uplink data transmission) per coherence block.


\textbf{Uplink channel estimation}:
Since each UE sends a distinct local update, a dedicated pilot assignment scheme (i.e. pilot sequences assigned for all UEs are pairwisely orthogonal) for channel estimation is considered. The orthogonality assumption simplifies the analysis and enables us to evaluate the performance of our proposed CFmMIMO-based multiple-FL system. The case of non-orthogonal pilots is left for future work.
Let $\tau_{dp}$ be the pilot length. Then, with a dedicated pilot assignment scheme, we need to have that  $\tau_{dp}\geq \sum_{n\in\NN} K_n$. The corresponding MMSE estimate $\bar{g}_{mk}$ of $g_{mn_k}$ is distributed according to $\CN(0,\bar{\sigma}_{mn_k}^2)$, where
$\bar{\sigma}_{mn_k}^2=\frac{\tau_{dp} \rho_{p} (\beta_{mn_k})^2} {\tau_{dp} \rho_{p} \beta_{mn_k} +1 }$ \cite{ngo17TWC}.

%
%


\textbf{Uplink payload data transmission}:
After sending the pilots for the channel estimation, each user will send its local update to the CPU. The signal transmitted from UE $k$ of group $n$  is given by $x_{u,{n_k}}\!=\!\sqrt{\rho_{u}\zeta_{n_k}}s_{u,{n_k}}$, 
where $s_{u,n_k}$, with $\EEE\{ |s_{u,n_k} |^2 \}=1$, is the associated symbol, $\rho_{u}$ is the maximum normalized transmit power at each UE, and $\zeta_{n}$ is the power control coefficient. The power control coefficients are chosen so that  the average transmit power is constrained, i.e., $\EEE\left\{|x_{u,n_k}|^2\right\}\leq \rho_u$, which  can be expressed through a per-UE power constraint as
\begin{align}\label{poweruupperbound}
\zeta_{n_k}\leq 1,\forall n\in\NN,n_k\in \K_n.
\end{align}
Using the signals received from all the UEs, the APs use the channel estimates to compute and send match-filtered signals to the CPU for detecting the UEs' message symbols. 
The achievable rate
$R_{u,{n_k}}(\ZETA)$ (bps) of UE $k$ in group $n$ is given by 
\eqref{Ruuni} \cite[Eq.~(27)]{ngo17TWC}, shown at the top of the next page, where $\ZETA \triangleq \{\zeta_{n_k}\}_{n\in\NN,k\in\K_n}$ 
\begin{figure*}[t!]
\begin{align}
\label{Ruuni}
&R_{u,n_k}(\ZETA)=\frac{\tau_c-\tau_{dp}}{\tau_c}B\log_2\Bigg(1+
\frac
{\rho_u\zeta_{n}\left(\sum_{m\in\MM}\bar{\sigma}_{mn_k}^2\right)^2}
{\rho_u\sum_{n'\in\NN}\sum_{\ell\in\K_{n'}}\zeta_{n'_\ell}\sum_{m\in\MM}\bar{\sigma}_{mn_k}^2\beta_{mn'_\ell}
+\sum_{m\in\MM}\bar{\sigma}_{mn_k}^2}
\Bigg),
\end{align}
\hrulefill
\vspace{-4mm}
\end{figure*}

\textbf{Uplink delay}:
Denote by $S_{u,n}$ (bits) the data size of the local training update of group $n$.
The transmission time (uplink delay) from UE $k$ of group $n$ to the APs is given by
\begin{align}\label{}
t_{u,n_k}(\ZETA) = \frac{S_{u,n}}{R_{u,n_k}(\ZETA)}.
\end{align}

\vspace{-1mm}
\begin{remark}
We want to emphasize that the achievable downlink and uplink rates given in \eqref{Rdmulti} and \eqref{Ruuni}, respectively, are obtained under the case that all users participate in the transmission. However, as we can see from the transmission protocol in Fig.~\ref{fig:time}, for a given time, some users may have finished their transmission, and hence, do not participate in the downlink or uplink transmission with other users at the same time. This will not cause any issue with our design since the rates \eqref{Rdmulti} and \eqref{Ruuni} are still achievable under this case.
\end{remark}
\vspace{-2mm}

\subsubsection{Step (S4)}
After receiving all the local updates, the CPU computes its global update. Since the CPU computational capability is far more powerful than that of the UEs, the delay of computing the global update is assumed as negligible. 
\vspace{-1mm}

\section{Problem Formulation and Solution}
\label{sec:PF}
\vspace{-1mm}
Reducing the learning time is one of the key targets of wireless communications-based FL systems. Thus, in this section, we propose to allocate the transmit powers and processing frequency to minimize the time of one FL iteration of every group.\footnote{In general, the objective should be the time of the whole FL process of every group. However, minimizing this time is very challenging and somehow impractical due to sophisticated synchronization requirements. In addition, reducing the learning time of each iteration will also lead to a reduction in  the learning time of the whole FL processing.}
%
As shown in Fig.~\ref{fig:time}, at the current iteration, every group needs to wait until all groups finish their iterations to start a new iteration. Therefore, the time of one FL iteration of every group is the longest delay caused by one UE in the network. Therefore, we formulate an optimization problem as follows:
\vspace{-1mm}
\begin{subequations}\label{Pmulti}
\begin{align}
\label{CFPmulti}
\!\!\!\!\!\underset{\ETA,\f,\ZETA}{\min} \,\,
&\max_{n\in\NN}\max_{n_k\in\K_n}\!
\big( t_{d,n_k}(\ETA) + t_{C,n_k}(f_{n_k}) + t_{u,n_k}(\ZETA) \big)
\\
\nonumber
\mathrm{s.t.}\,\,
&
\eqref{powerdupperbound}, \eqref{poweruupperbound}
\\
\label{powerlowerbound}
& 0\leq \eta_{mn}, 0\leq \zeta_{n_k}, \forall m,n,n_k
\\
\label{fbound}
& 0 \leq f_{n_k} \leq f_{\max}, \forall n,n_k
\\
\label{syncbound}
& \max_{n\in\NN}\max_{n_k\in\K_n}\!\! t_{d,n_k}
\!\leq\! \min_{n\in\NN}\min_{n_k\in\K_n} \big(t_{d,n_k} \!+\! t_{C,n_k}\big),
\end{align}
\end{subequations}
\noindent
where $\f\triangleq\{f_{n_k}\}_{n\in\NN,n_k\in\K_n}$. Here, \eqref{syncbound} is introduced to ensure that every UE sends its local update during the uplink mode of the APs. In particular, it models the scenario that the UE, which finishes its downlink transmission and local computation the earliest, starts its uplink transmission after the UE, which finishes its downlink transmission the latest, starts its local computation (as seen in Fig.~\ref{fig:time}(a)). 

The problem \eqref{Pmulti} can be rewritten in an epigraph form as
\begin{subequations}\label{Pmultiepi}
\begin{align}
\label{CF:shortP:epi}
\!\!\!\!\underset{\x}{\min} \,\,
& a
\\
\mathrm{s.t.}\,\,
\nonumber
&
\eqref{powerdupperbound}, \eqref{poweruupperbound}, \eqref{powerlowerbound}
\\
\label{Rdmultilowerbound}
& r_{d,n_k}\leq R_{d,n_k} (\ETA), \forall n,n_k
\\
\label{Rumultilowerbound}
& r_{u,n_k}\leq R_{u,n_k} (\ZETA), \forall n,n_k
\\
\label{rlowerbound}
&  0 \leq r_{d,n_k}, 0 \leq r_{u,n_k}, \forall  n,n_k
\\
\label{tbound}
& \frac{S_{d,n}}{r_{d,n_k}} + \frac{LD_nc_{n_k}}{f_{n_k}} + \frac{S_{u,n}}{r_{u,n_k}} \leq a, \forall n, n_k
\\
\label{syncbound2a}
& b \leq q
\\
\label{syncbound2b}
& \frac{S_{d,n}}{r_{d,n_k}} \leq b, \forall n,n_k
\\
\label{syncbound2c}
& q \leq q_{1,n_k} + q_{2,n_k}, \forall n,n_k
\\
\label{syncbound2d}
& 0 \leq q_{1,n_k}, 0 \leq q_{2,n_k}, \forall n,n_k
\\
\label{syncbound2e}
& q_{1,n_k} \leq \frac{S_{d,n}}{r_{d,n_k}}, \forall n,n_k
\\
\label{syncbound2f}
& q_{2,n_k} \leq \frac{LD_nc_{n_k}}{f_{n_k}}, \forall n,n_k,
\end{align}
\end{subequations}
where $\x \triangleq \{\ETA,\f,\ZETA,\rr_d,\rr_u,a,b,q,\q_1,\q_2\}$, $\rr_d=\{r_{d,n_k}\}$, $\rr_u=\{r_{u,n_k}\}$, 
$\q_1=\{q_{1,n_k}\}$, 
$\q_2=\{q_{2,n_k}\}, \forall n\in\NN,n_k\in\K_n$, 
and $\rr_d,\rr_u,a,b,q,\q_1,\q_2$ are additional variables. Here, 
\eqref{syncbound2a}--\eqref{syncbound2f} follow from \eqref{syncbound}.
If we let
$\vv \triangleq \{v_{mn}\}_{m\in\MM,n\in\NN}$ and $\uu\triangleq \{u_{n_k}\}_{n\in\NN,n_k\in\K_n}$ with
\begin{align} \label{variabletransform}
v_{mn}\triangleq \eta_{mn}^{1/2}, \,\, u_{n_k} \triangleq \zeta_{n_k}^{1/2}, \forall m,n,n_k,
\end{align}
the problem \eqref{Pmultiepi} is, then, equivalent to 
\vspace{-1mm}
\begin{subequations}\label{Pmultiepiequi}
\begin{align}
\label{CF:shortP:equi}
\underset{\widehat{\x}}{\min} \,\,
& a
\\
\nonumber
\mathrm{s.t.}\,\,
&\eqref{rlowerbound}-\eqref{syncbound2f}
\\
\label{Rdmultilowerbound2}
& r_{d,n_k} \leq R_{d,n_k}(\vv), \forall n,n_k
\\
\label{Rumultilowerbound2}
& r_{u,n_k} \leq R_{u,n_k}(\uu), \forall n,n_k
\\
\label{powerdupperbound2}
& \sum_{n\in\NN} \sum_{k\in\K_n} \sigma_{mn_k}^2v_{mn}^2 \leq 1, \forall m
\\
\label{poweruupperbound2}
& u_{n_k}^2\leq 1, \forall n,n_k
\\
\label{powerlowerbound2}
& 0\leq v_{mn}, 0\leq u_{n_k},\forall m,n,n_k,
\end{align}
\end{subequations}
\noindent
where $\widehat{\x}\triangleq\{\x,\vv,\uu\}\setminus\{\ETA,\ZETA\}$; \eqref{Rdmultilowerbound2}, \eqref{Rumultilowerbound2} follow \eqref{Rdmultilowerbound}, \eqref{Rumultilowerbound}, while \eqref{powerdupperbound2}--\eqref{poweruupperbound2} follow \eqref{powerdupperbound}, \eqref{poweruupperbound}, \eqref{powerlowerbound}.
The functions in the right-hand sides of the nonconvex constraints \eqref{syncbound2e} and \eqref{syncbound2f} have the following concave lower bounds:
\begin{align}\label{qupperbound}
&h_{1,n_k}(r_{d,n_k})\triangleq
S_{d,n}\Big(\frac{2}{r_{d,n_k}^{(i)}} - \frac{r_{d,n_k}}{(r_{d,n_k}^{(i)})^2}\Big) \leq \frac{S_{d,n}}{r_{d,n_k}},
\\
&h_{2,n_k}(f_{n_k})\triangleq LD_nc_{n_k} \Big(\frac{2}{f_{n_k}^{(i)}} - \frac{f_{n_k}}{(f_{n_k}^{(i)})^2}\Big) \leq  \frac{LD_nc_{n_k}}{f_{n_k}}.
\end{align}
For the nonconvex constraints \eqref{Rdmultilowerbound2} and \eqref{Rumultilowerbound2}, $R_{d,n_k}(\vv)$ has a concave lower bound $\widetilde{R}_{d,n_k}(\vv)$, which is given by \cite{vu20TWC}
\begin{align}\label{Rdconcave}
\nonumber
&\widetilde{R}_{d,n_k}(\vv) \triangleq 
\frac{\tau_c-\tau_{cp}}{\tau_c \log 2} B \Big[ \log \Big(1 + \frac{(\Upsilon_{n_k}^{(i)})^2}  {\Pi_{n_k}^{(i)}} \Big)
-\frac{(\Upsilon_{n_k}^{(i)})^2}{\Pi_{n_k}^{(i)}}
\\
&+2\frac{\Upsilon_{n_k}^{(i)}\Upsilon_{n_k}}{\Pi_{n_k}^{(i)}}
-\frac{(\Upsilon_{n_k}^{(i)})^2(\Upsilon_{n_k}^2+\Pi_{n_k})}{\Pi_{n_k}^{(i)}((\Upsilon_{n_k}^{(i)})^2+\Pi_{n_k}^{(i)})}\Big]
\leq R_{d,n_k}(\vv),
\end{align}
\noindent
where $\Pi_{n_k}(\vv) = \rho_d\sum_{n'\in\NN} \sum_{m\in\MM} v_{mn'}^2 \beta_{mn_k} \big( \sum_{n'_\ell\in\K_n} \\ \hat{\sigma}_{mn'_\ell} \big)^2  + 1$, and $\Upsilon_{n_k}(\{v_{mn}\}_{m\in\MM}) = \sqrt{\rho_d}
\sum_{n_\ell\in\K_n} \sum_{m\in\MM}\\ v_{mn} \hat{\sigma}_{mn_k} \hat{\sigma}_{mn_\ell}$. Similarly, $R_{u,n_k}(\uu)$ has a concave lower bound $\widetilde{R}_{u,n_k}(\uu)$, which is expressed as
\begin{align}\label{Ruconcave}
\nonumber
&\widetilde{R}_{u,n_k}(\uu) \triangleq
\frac{\tau_c-\tau_{dp}}{\tau_c \log 2} B \Big[ \log\Big(1+\frac{(\Psi_{n_k}^{(i)})^2}{\Xi_{n_k}^{(i)}}\Big)
-\frac{(\Psi_{n_k}^{(i)})^2}{\Xi_{n_k}^{(i)}}
\\
&+ 2\frac{\Psi_{n_k}^{(i)}\Psi_{n_k}}{\Xi_{n_k}^{(i)}}
- \frac{(\Psi_{n_k}^{(i)})^2(\Psi_{n_k}^2+\Xi_{n_k})}{\Xi_{n_k}^{(i)}((\Psi_{n_k}^{(i)})^2+\Xi_{n_k}^{(i)})} \Big]
\leq R_{u,n_k}(\uu),
\end{align}
\noindent
where $\Xi_{n_k}(\uu)\!\! =\!\!
\rho_u\sum_{n'\in\NN}  \sum_{n'_\ell\in\K_{n'}} u_{n'_\ell}^2 \sum_{m\in\MM} \bar{\sigma}_{mn_k}^2 \beta_{mn'_\ell}
+\sum_{m\in\MM}\bar{\sigma}_{mn_k}^2$, and $\Psi_{n_k}(u_{n_k}) =
\sqrt{\rho_u} u_{n_k}(\sum_{m\in\MM}\bar{\sigma}_{mn_k}^2)$.

As such, \eqref{syncbound2e}, \eqref{syncbound2f}, \eqref{Rdmultilowerbound2}, and \eqref{Rumultilowerbound2} can be approximated respectively by the following convex constraints:
\begin{align}
\label{syncbound2eapprox}
&q_{1,n_k} \leq h_{1,n_k}(r_{d,n_k}), \forall n,n_k,
\\
\label{syncbound2fapprox}
&q_{2,n_k} \leq h_{2,n_k}(f_{n_k}) , \forall n,n_k,
\\
\label{Rdmultilowbound2approx}
&r_{d,n_k}  \leq \widetilde{R}_{d,n_k}(\vv) , \forall n,n_k,
\\
\label{Rumultilowbound2approx}
&r_{u,n_k}  \leq \widetilde{R}_{u,n_k}(\uu) , \forall n,n_k.
\end{align}
\noindent
At iteration $(i+1)$, for a given point $\widehat{\x}^{(i)}$, problem \eqref{Pmultiepiequi} (and hence, \eqref{Pmultiepi}) can finally be approximated by the following convex problem:
\vspace{-1mm}
\begin{align}\label{Pmultiepiequiapprox}
\underset{\widehat{\x}\in\widetilde{\FF}}{\min} \,\,
a
\end{align}
where $\widetilde{\FF}\triangleq\!\{
\eqref{rlowerbound}-\eqref{syncbound2d},
\eqref{powerdupperbound2} - \eqref{powerlowerbound2}, 
\eqref{syncbound2eapprox} - 
\eqref{Rumultilowbound2approx}
\}$ is a convex feasible set.

In Algorithm~\ref{alg}, we outline the main steps to solve problem \eqref{Pmulti}.
Let $\FF\triangleq\{
\eqref{powerdupperbound}, \eqref{poweruupperbound},
\eqref{powerlowerbound} - 
\eqref{syncbound}\}$ be the feasible set of \eqref{Pmulti}.
Starting from a random point $\widehat{\x}\in\FF$, we solve \eqref{Pmultiepiequiapprox} to obtain its optimal solution $\widehat{\x}^*$, and use $\widehat{\x}^*$ as an initial point in the next iteration. The algorithm terminates when an accuracy level of $\varepsilon$ is reached.
In the case when $\widetilde{\FF}$ satisfies Slater's constraint qualification condition, Algorithm~\ref{alg} will converge to a Karush-Kuhn-Tucker (KKT) solution of \eqref{Pmultiepiequi} \cite[Theorem 1]{Marks78OR}. In contrast, Algorithm~\ref{alg} will converge to a Fritz John (FJ) solution of \eqref{Pmultiepiequi}. By using the variable transformation \eqref{variabletransform}, it can be seen that the KKT (resp. FJ) solution of \eqref{Pmultiepiequi} satisfies the KKT (resp. FJ) conditions of \eqref{Pmultiepi} and \eqref{Pmulti}.

\begin{algorithm}[!t]
\caption{Solving problem \eqref{Pmulti}}
\begin{algorithmic}[1]\label{alg}
\STATE \textbf{Initialize}: Set $i\!=\!0$ and choose a random point $\widehat{\x}^{(0)}\!\in\!\FF$.
\REPEAT
\STATE Update $i=i+1$
\STATE Solving \eqref{Pmultiepiequiapprox} to get its optimal solution $\widehat{\x}^*$
\STATE Update $\widehat{\x}^{(i)}=\widehat{\x}^*$
\UNTIL{convergence}
\end{algorithmic}
\vspace{-1mm}
\textbf{Output}: $(\ETA^*,\ZETA^*,\f^*)$
\end{algorithm}
\vspace{-1mm}

\section{Numerical Examples}
\vspace{-1mm}
\label{sec:sim}
\subsection{Network Setup and Parameter Setting}
\vspace{-1mm}
Consider a CFmMIMO network, where the APs and UEs are randomly located in a square of $D\times D$ km$^2$. The distances between adjacent APs are at least $50$ m. 
We set $\tau_c\!=\!200$ samples.
The large-scale fading coefficients, i.e., $\beta_{mn_k}$, are modeled in the same manner as \cite[Eqs. (37), (38)]{emil20TWC}.
For ease of presentation, all groups have the same number of UEs, i.e., $K_n =K, \forall n$. The total number of UEs is thus $NK$.
We choose $\tau_{cp} \!=\! N$, $\tau_{dp} \!=\! NK$, $S_d\!=\!S_u\!=\!5$ MB, noise power $\sigma_0^2\!=\!-92$ dBm, $L=50$, $f_{\max}=3 \times 10^9$ cycles/s, $D_n = 5\times 10^6$ samples, $c_{n_k} = 20$ cycles/samples \cite{tran19INFOCOM}, for all $n,n_k$, $\alpha=2\times 10^{-29}$. 
Let $\tilde{\rho}_d\!=\!1$ W, $\tilde{\rho}_u\!=\!0.2$ W and $\tilde{\rho}_p\!=\!0.2$ W be the maximum transmit power of the APs, UEs and uplink pilot sequences, respectively. The maximum transmit powers $\rho_d$, $\rho_u$ and $\rho_p$ are normalized by the noise power.  
\vspace{-1mm}

\subsection{Results and Discussions}
\vspace{-1mm}
To evaluate the effectiveness of our proposed scheme (\textbf{Joint\_OPT\_CF}), we consider the following baseline schemes
\begin{itemize}
    \item \textbf{Separate\_OPT\_CF}: Processing frequencies $\f$ are optimized first, given the same downlink power to all groups, i.e., $\eta_{mn}\sum_{n_k\in \K_n}\sigma_{mn_k}^2 \!=\! 1/N, \forall n$ and the transmitted power of each UE is $\eta_{n_k}=1, \forall n, n_k$. The optimized values of $\f$ are then used as the input for optimizing $\ETA$ and $\ZETA$. 
    \item \textbf{Joint\_OPT\_Co}: For the same setting and number of antennas, we apply our Algorithm.~\ref{alg} for colocated massive MIMO (mMIMO), which is a special case of CFmMIMO where all the APs are  colocated at a base station in the center of the considered square. 
\end{itemize} 
\vspace{-0mm}

\begin{figure}[t!]
  \centering
  {\includegraphics[width=0.36\textwidth]{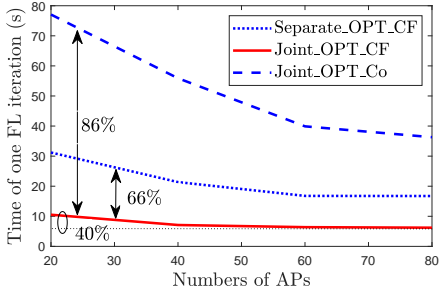}\label{fig:a}}
  \caption{Comparison among the proposed approach and baselines ($K = 8$ (users per group), $N\!=\!3$ groups, and $D=0.5$ km).}
  \label{Fig:sim1}
  \vspace{-5mm}
\end{figure}

\begin{figure}[t!]
  \centering
  {\includegraphics[width=0.36\textwidth]{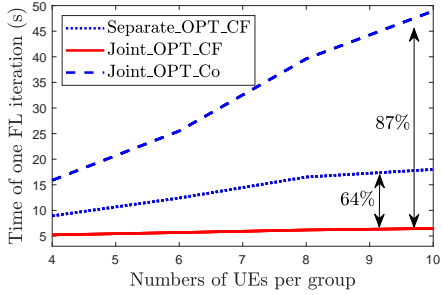}\label{fig:b}}
  \vspace{-0mm}
  \caption{Comparison among the proposed approach and baselines ($M = 80$ (antennas), $N\!=\!3$ groups, $D=0.5$ km).}
  \label{Fig:sim2}
  \vspace{-0mm}
\end{figure}

Figs.~\ref{Fig:sim1} and~\ref{Fig:sim2} compare the time of one FL iteration with the considered schemes. As seen, Algorithm~\ref{alg} gives the best performance. Specifically, compared to Separate\_OPT\_CF, the time reductions by Algorithm~\ref{alg} are up to $66\%$  with $M=20$, $K=8$, and up to $64\%$ with $M=80$, $K=10$. The figures not only demonstrate the noticeable advantage of a joint allocation of power and processing frequency, but also show the benefit of using massive MIMO to support FL. Thanks to massive MIMO technology, the data rates of each UE increases when the number of antennas (i.e., APs) increases, leading to a decrease of $40\%$ in the execution time of one FL iteration as shown in Fig.~\ref{Fig:sim1}. 

Figs.~\ref{Fig:sim1} and~\ref{Fig:sim2} also show that CFmMIMO significantly outperforms colocated mMIMO. In particular, the reduction in the time of one FL iteration is by up to $87\%$ with  $M=80$, $K=10$. This is reasonable because CFmMIMO, with antennas distributed over a geographic area, is likely to suffer less from UEs with unfavorable links than colocated mMIMO. Higher coverage probability and lower training time are thus expected.
\vspace{-2mm}
\section{Conclusion}
\label{sec:con}
\vspace{-0mm}
This work has proposed a novel scheme with CFmMIMO as a solution for future wireless networks to support multiple FL groups. Using successive convex approximation techniques, we have also successfully proposed an algorithm to allocate power and processing frequency in order to optimally reduce the training time of each FL iteration. Numerical results showed that our proposed algorithm significantly reduces the time of each FL iteration over the baseline schemes. They also confirmed that with the same maximum ratio technique to support multiple FL groups, CFmMIMO is a better choice than colocated mMIMO. 
\vspace{-0mm}

\ifCLASSOPTIONcaptionsoff
  \newpage
\fi

\section*{Acknowledgment}
The work of T.~T.~Vu and H.~Q.~Ngo was supported by the U.K. Research and Innovation Future Leaders Fellowships under Grant MR/S017666/1. The work of M.~Matthaiou was supported by a research grant from the Department for the Economy Northern Ireland under the US-Ireland R\&D Partnership Programme and by the European Research Council (ERC) under the European Union’s Horizon 2020 research and innovation programme (grant agreement No.~101001331).

\begin{spacing}{0.93}
\bibliographystyle{IEEEtran}
\bibliography{IEEEabrv,newidea2021}
\end{spacing}

\end{document}